# Statistical mechanics of damage phenomena

Short title: Statistical…

S.G. Abaimov[1]

[1]University of Western Ontario, London, ON, N6A 5B7, Canada

E-mail: sabaimov@uwo.ca

**Abstract:** This paper applies the formalism of classical, Gibbs-Boltzmann statistical mechanics to the phenomenon of non-thermal damage. As an example, a non-thermal fiber-bundle model with the global uniform (meanfield) load sharing is considered. Stochastic topological behavior in the system is described in terms of an effective temperature parameter thermalizing the system. An equation of state and a topological analog of the energy-balance equation are obtained. The formalism of the free energy potential is developed, and the nature of the first order phase transition and spinodal is demonstrated.

**PACS.** 62.20.M- Structural failure of materials - 89.75.-k Complex systems - 05. Statistical physics, thermodynamics, and nonlinear dynamical systems

**Used abbreviations:**

SM – statistical mechanics;

ASM – in a complete analogy with statistical mechanics;

ESC – external stochastic constraint (*e.g.,* temperature dictated by the medium in a canonical ensemble);

EBC – external boundary constraint (*e.g.,* constant volume or pressure prescribed to the system);

FBM – fiber-bundle model.

**1 Introduction**

Damage as a complex phenomenon has been studied by many authors and a survey of recent developments in damage mechanics can be found in [1-3]. Although temperature as a parameter could be included in consideration the models of damage usually don't require its presence. Instead, the damage phenomenon is described by the topological stochasticity of fracture surfaces. Stochastic 'geometric' behavior of the system carries on the role of thermal fluctuations.

A natural question arises: Can we describe the topological fluctuations in the system in terms of an effective temperature parameter? If the temperature is naturally absent but the stochastic behavior carries on its role the question above can be reformulated as: Can the formalism of statistical mechanics (further SM) be extended to non-thermal damage phenomena?

Many attempts [4-10] have been made to answer these questions. One of the most brilliant approaches was suggested by Smalley *et al.* [11]. To model damage they actually used Dyson's hierarchical model developed in SM [see 12 and further citations in the literature]. In spite of the great interest the application of SM to damage phenomena is still an unresolved question.

This paper develops a rigorous, systematic approach to the question. It could be expected that the direct applications of the energy concepts of SM to topological phenomena are neither correct nor possible. Instead, the basic variables of statistical mechanics like temperature, energy, and free energy potential should be mapped on their topological analogs. Therefore we should start from the basic principles and follow a step-by-step comparison between SM and non-thermal damage phenomenon. In this way we could find the topological analogs of the equation of state, energy-balance equation, and free energy potential. The knowledge of the free energy potential would give us a complete knowledge of the system. Particularly, the behavior of possible phase transitions in the system would be classified by it.

The nature of damage phenomenon is assumed to be represented by the stochasticity of its topological occurrence. The equilibrium properties of this stochastic behavior are assumed to be prescribed as an external stochastic constraint (further ESC). In other words, the equilibrium probability distributions are dictated by *a priori* assigned model stochasticity. The analogy with statistical mechanics describes this ESC in terms of the effective temperature parameter thermalizing the system. In other words, stochastic fluctuations are mapped on thermal fluctuations of SM and prescribed ESC is mapped on the ESC of the classic canonical ensemble (temperature dictated by the external medium). And the main purpose of the paper remains to illustrate a complete similarity between SM and damage mechanics. To emphasize this the abbreviation 'ASM' will be used when the results will appear to be in a complete analogy with SM.

**2 Model**

Damage is a complex phenomenon. It can be associated with local and non-local load sharing, brittle and ductile behavior. It can emerge both in one-dimensional and three-dimensional systems, leading in the last case to three-dimensional stress patterns of crack formation. The basic principles of damage are often completely disguised by the secondary side effects of its occurrence.

Therefore, to develop a theory of damage as a SM phenomenon, it is reasonable to consider initially a simple model. So, in SM the Van der Waals' meanfield model is usually used to illustrate the behavior of gas-liquid systems. In the case of magnetic systems an analogy is the meanfield Ising model with the infinite range of interactions. The basic principles of SM and phase-transition behavior can be illustrated by these simple models. Further models corresponding to real systems can be constructed as more accurate and more complex improvements.

In this paper we will investigate the meanfield approximation of damage phenomena. As a model we consider a static (deterministic, quenched) fiber-bundle model (further FBM) with the uniform global (meanfield) load sharing. The term 'static' as an opposite to the term 'stochastic' is used to specify that each fiber has *a priori* assigned strength threshold $s$ which does not change during the model evolution, and this fiber can fail only when its stress $\sigma_F$ exceeds its strength $s$. All fibers have predefined strengths, distributed with *a priori* specified probability density function $p_s(s)$. The cumulative distribution function $P_s(\sigma_F) = \int_0^{\sigma_F} p_s(s)ds$ is the probability for a fiber to have been already broken if its stress is supposed to be $\sigma_F$.

We assume that there is an ensemble of identical systems. The systems are non-thermodynamic and thermal fluctuations are naturally absent in them. Instead, each system in the ensemble realizes some strength distribution over its fibers. This introduces stochastic topological fluctuations. The prescribed distribution $P_s(s)$ works as an ESC introduced into the system as a model input. This ESC is similar to the temperature prescribed in a canonical ensemble. The external medium dictates the equilibrium distribution of probabilities for the stochastic behavior but the system actually can realize itself in a non-equilibrium state with any other probability distribution (*i.e.,* it can fluctuate from the equilibrium with the ESC). *E.g.,* in the thermal canonical ensemble a system can consume any amount of energy and only the equilibrium value is prescribed by the temperature. Similar, we assume that a system in the ensemble can realize an arbitrary distribution of fiber strengths and realizes $P_s(s)$ only in the equilibrium with the ESC. The model is deterministic (static) in the sense that for each particular system the realization of strength distribution is assigned *a priori* and does not change during the system evolution. Therefore for the same realization of strength distribution the system follows during its evolution exactly the same deterministic trajectory. However, the model is stochastic in the sense that *a priori* assigned sample distribution of a system is prescribed stochastically and only on the ensemble average should correspond to $P_s(s)$.

We assume that the number of fibers in the model $N$ is constant and infinite in the thermodynamic limit. Intact fibers carry all the same strain $\varepsilon_F$ which is identically equal to the strain $\varepsilon$ of the whole model as a 'black box': $\varepsilon_F \equiv \varepsilon$. The stress of each intact fiber is assumed to have a linear elastic dependence on the strain till the fiber failure: $\sigma_F = E\varepsilon_F$, where $E$ is the Young modulus, which is assumed to be the same for all fibers. This introduces the concept of non-linear stress-strain dependence for the total model although each fiber behaves elastically till its failure.

Order parameters in gas-liquid systems are densities of phases; in magnetic systems they are magnetizations of phases. Similarly to the SM approach we define different phases of damage as phases with different fractions of broken fibers. To do so it is necessary to introduce a damage parameter $D$ as a fraction of broken fibers. If the total number of fibers in the model is $N$ then the number of broken fibers is $DN$ and the number of intact fibers is $(1-D)N$. The damage parameter $D$ plays here a role of the order parameter distinguishing phases.

For all possible states of the system (equilibrium or not) we will assume that the total external force acting on the system is balanced by the response force of all fibers – transient processes with the discontinuity of boundary force are not considered in this paper. Then if we, as an external observer, were to look at the model as at a 'black box' we would not to know how many fibers are broken inside and we would see that the external force $F$ is applied to $N$ fibers, creating inform 'virtual' stress $\sigma$ at the model surface. The actual stress in fibers is $1 / (1 - D)$ times higher due to the fiber failure and stress redistribution. Therefore, the virtual stress of the model (how an external observer sees it) is $\sigma = (1 - D)\sigma_F$ or

$$\sigma = (1 - D)E\varepsilon. \tag{1}$$

Under the condition of fixed number of particles liquid-gas and magnetic systems have two field parameters outside of coexistence region: these could be temperature and volume/magnetization, or temperature and pressure/magnetic field, or any other pair of independent variables. In the case of FBM strain $\varepsilon$ or stress $\sigma$ could be chosen to be a field parameter. However, in contrast to other systems, only one of these parameters is independent. 'Geometric' constraint (1) and an equation of state make two of three parameters $D$, $\varepsilon$, $\sigma$ dependent. Therefore we would preliminary expect that an analog of the energy-balance equation for the FBM must have one term less.

In SM a system has a phase space of microstates along which it wonders jumping from one microstate to another in accordance with the assigned probabilities of each microstate. In general, the system can reach any macrostate of all possible, and only the long waiting time makes the non-equilibrium states unobservable. Another formulation is a concept of ensemble of identical systems. An observer can simultaneously watch different microstates as realizations of different copies of the initial system. An ergodic hypothesis claims that these two different approaches are equivalent: in the ensemble at a given time the observer sees each state realized with the same probability as it would be in the case of observations of only one system for the long time.

However, in the case of the FBM without healing (which is under the consideration) we have only one possibility – an ensemble of identical systems. Each realization of the model has *a priori* prescribed strength distribution which does not change during the system evolution. Therefore, for the given value of constant strain or stress the system has failed in one particular way and will never visit other microstates. So, only ensemble analogy with SM is possible for the FBM.

**3 Definition of a microstate, macrostate, and equilibrium state**

In the case of gas-liquid systems microstates are defined as cells in the phase space of the system. In the case of magnetic systems microstates are defined as different realizations of "up" and "down" spins on a lattice: ↑↑↓↓↑. We can define microstates for the FBM in an analogous way as different realizations of broken and intact fibers. So, for the FBM with $N = 3$ fibers all possible microstates are |||, ||x, |x|, x||, |xx, x|x, xx|, and xxx where symbol '|' denotes an intact fiber while symbol 'x' – a broken fiber. Further in the paper index $n$ will be used to enumerate all possible microstates. A similarity with the Ising model should be mentioned here because each spin or fiber is fixed at a particular location of the lattice and all fibers are macroscopic objects. Therefore in contrast to particles, they should not be considered as identical (indistinguishable) objects.

For the specified damage $D$ and for the external boundary constraints (further EBC) $\varepsilon$ = const or $\sigma$ = const each microstate has probability

$$w_n^{equil}(D) \equiv w^{equil}(D) = (1 - P_s(E\varepsilon))^{(1-D)N} (P_s(E\varepsilon))^{DN} \text{ for } \varepsilon = \text{const or} \qquad (2a)$$

$$w_n^{equil}(D) \equiv w^{equil}(D) = \left(1 - P_s\left(\frac{\sigma}{1-D}\right)\right)^{(1-D)N} \left(P_s\left(\frac{\sigma}{1-D}\right)\right)^{DN} \quad \text{for } \sigma = \text{const} \tag{2b}$$

as the probability that $(1 - D)N$ fibers are intact and $DN$ fibers are broken. This probability $w_n^{equil}$ is dictated by the ESC $P_s(s)$. Again, this ESC is similar to the temperature prescribed in a canonical ensemble. The external medium dictates the equilibrium distribution of probabilities but the system actually can realize itself in a non-equilibrium state with any other probability distribution $w_n$. Only the equilibrium state is dictated by the ESC therefore we used abbreviation 'equil' to emphasize that this probability distribution corresponds to the equilibrium with the ESC.

In general, a macrostate may be defined as a union of all possible microstates realized with the specified probabilities. However, in this paper a simpler definition will be used when these probabilities equal only zero or unity. In other words, further in this paper as to a macrostate we will refer to a subset of microstates chosen by a particular property. For example, the definition of a macrostate in a thermal canonical ensemble is a subset of all microstates with the specified energy. For the FBM we will presumably use the definition of a macrostate as a subset of all microstates with the specified fraction of broken fibers, *i.e.*, with the specified damage $D$. If another definition of a macrostate shall be utilized it will be clearly specified.

All microstates corresponding to the macrostate $D$ have the same probability (2) and the number of these microstates is given by the combinatorial choice of $DN$ broken fibers among $N$ fibers

$$N_D^{macro} = \frac{N!}{(DN)!((1-D)N)!} \approx_{\ln} \frac{1}{D^{DN}(1-D)^{(1-D)N}} \tag{3}$$

where the symbol "$\approx_{\ln}$" means that in the thermodynamic limit $N \to +\infty$ all power-law multipliers are neglected in comparison with the exponential dependence on $N$. Everywhere further the symbol "$\approx_{\ln}$" will mean the accuracy of exponential dependence neglecting all power-law dependences. For the logarithm of such equations we will use the symbol "$\approx$".

To find the total probability of the macrostate $D$ we need to multiply the probability of each microstate (2) by the total number of these microstates (3) corresponding to the given macrostate

$$W_{macro\,D}^{equil}(D) = \sum_{n=1}^{N_D^{macro}} w_n^{equil}(D) = N_D^{macro} w^{equil}(D) \approx_{\ln} \left(\frac{1-P_s(E\varepsilon)}{1-D}\right)^{(1-D)N} \left(\frac{P_s(E\varepsilon)}{D}\right)^{DN} \tag{4a}$$

for the EBC $\varepsilon$ = const and

$$W_{macro\,D}^{equil}(D) \approx_{\ln} \left(\frac{1-P_s\left(\frac{\sigma}{1-D}\right)}{1-D}\right)^{(1-D)N} \left(\frac{P_s\left(\frac{\sigma}{1-D}\right)}{D}\right)^{DN} \tag{4b}$$

for the EBC $\sigma$ = const. This is the probability for the macrostate $D$ to be observed in equilibrium with the ESC.

For the equilibrium we will use two different definitions which are often confused in the literature. The ESC is assumed to prescribe the equilibrium probability distribution $w_n^{equil}$ for all microstates. Therefore, the equilibrium with the ESC could be identified with a system which can realize itself on all microstates with equilibrium probabilities: $w_n = w_n^{equil}$. In other words, all microstates are possible but their probabilities are dictated by the ESC. As an example we could think of the equilibrium in a thermal canonical ensemble where all microstates with all energies are possible but their probabilities are dictated by Boltzmann distribution. The superscript 'equil' will be used for this definition. Then the value of any quantity *f* in equilibrium by definition is $f^{equil} \equiv \sum_n w_n^{equil} f_n$.

In contrast, another definition is the equilibrium (most probable) macrostate, *i.e.*, a system that can realize itself only on (that is isolated on) a subset of microstates corresponding to the most probable macrostate. This is the macrostate which gives the main contribution to the partition function. As an example we could think of the equilibrium macrostate of the thermal canonical ensemble where we count only those microstates whose energies equal to the equilibrium value $E_0 = Nk_BT/2$. To distinguish this case the subscript '$_0$' will be used.

**4 Why we cannot directly apply the SM formalism**

The FBM is not a thermal system and it has topological fluctuation instead of thermal fluctuations. Temperature is naturally absent in this system. Therefore the methods of SM cannot be applied directly and have to be developed independently. For example, under constant stress and constant temperature conditions it would not be correct to say that the free energy potential of the system (whose minimization gives an equilibrium state) is the Gibbs free energy *G* defined in the traditional way $G = E - TS - V\sigma\varepsilon$ where *E* is the energy of the system. The FBM model does not have temperature or energy-balance equation. Therefore SM is not applicable here directly.

Another incorrect method would be the minimization of entropy as it is done in SM. One of definitions of entropy is given by

$$S \equiv -<\ln w_n> \equiv -\sum_n w_n \ln w_n \qquad (5)$$

where $w_n$ is the probability of microstate $n$ and the sum goes over all microstates (here and further Boltzmann constant is omitted for the formulas of SM and is absent naturally for the FBM formulas). For the FBM this formula can be rewritten as

$$S = -\sum_D N_D^{macro} w(D) \ln w(D) \qquad (6)$$

where we have accumulated the summation over all microstates corresponding to the specified $D$ by the number of these microstates $N_D^{macro}$ due to the fact that all these microstates have equal probabilities $w(D)$ given by Eq. (2).

For an isolated system the negative entropy is the free energy potential and the equilibrium state can be found by the maximization of entropy over all possible values of microstate probabilities $w_n$. The method of Lagrange multipliers is used with the single EBC prescribed for the system that the sum of probabilities of all microstates must be equal to unity: $\sum_n w_n = 1$. Then the maximization gives equal probabilities $w_n^{equil}$ for all microstates.

For a canonical ensemble a correct equilibrium state is usually found by adding an additional EBC prescribed to the system: the mean energy must be equal to the energy, dictated by the medium: $\sum_n E_n w_n = E_0$. It should be said however that this condition is irrelevant for the canonical ensemble (because nothing restricts the system to consume any energy) and is *in situ* only an artificial trick to obtain the result by the maximization of entropy. With this trick the maximization of entropy gives the correct Gibbs distribution $w_n^{equil}$. However, a better way to obtain this result would be to say that for the canonical ensemble (constant temperature and constant volume) the free energy potential is not the entropy but the Helmholtz energy $A = E - TS = \sum_n E_n w_n + T \sum_n w_n \ln w_n$. Minimization of this quantity by the method of Lagrange multipliers with the single relevant to the canonical ensemble EBC $\sum_n w_n = 1$ (without artificially introduced EBC $\sum_n E_n w_n = E_0$) gives the correct result – the Gibbs distribution $w_n^{equil}$ [see, e.g. 13].

For the FBM this method mustn't be used because the maximization of microstate probabilities $w_n$ must include the ESC $P_s(s)$ as an additional EBC. In other words, we must minimize the free energy over probabilities $w_n$ with the constraints $w_n \equiv w_n^{equil}$ where $w_n^{equil}$ is given by Eq. (2). Therefore, there is no variability for $w_n$ and such maximization wouldn't have any sense.

## 5 Equation of state for the constant strain $\varepsilon$ = const as an EBC

$W_{macro\,D}^{equil}(D) = N_D^{macro} w^{equil}(D)$ is the probability distribution of macrostates in equilibrium with the ESC. To obtain the averaged values of measurable quantities in this state it is necessary to maximize $W_{macro\,D}^{equil}(D)$ over all possible values of $D$: $0 \leq D \leq 1$. Both functions $N_D^{macro}$ and $w^{equil}(D)$ depend exponentially on $N$ which is infinite in the thermodynamic limit. Therefore these functions have very rapid change with the change of $D$ and the maximum of $W_{macro\,D}^{equil}(D)$ is very sharp (ASM). To find the maximum it is necessary to find when derivatives $\frac{dW_{macro\,D}^{equil}}{dD}(D)$ or $\frac{d\ln W_{macro\,D}^{equil}}{dD}(D)$ equal to zero. After simple algebra we obtain that the equilibrium value of $D_0$ for the given $\varepsilon$ is

$$D_0 = P_s(E\varepsilon) \tag{7}$$

as it could be expected because the equilibrium damage must be determined by the number of failed fibers. An example of the dependence of $W_{macro\,D}^{equil}(D)$ on $D$ and the maximum $D_0$ are given in Fig. 4a for $N = 100$. Eq. (7) is **the equation of the equilibrium state** in contrast to Eq. (1) which could be named as an equation of the 'geometric' constraint.

It is easy to find that the second derivative of the function $\ln W_{macro\,D}^{equil}(D)$ at the point of maximum (7) equals $\frac{-N}{P_s(E\varepsilon)(1-P_s(E\varepsilon))}$. The second derivative is negative therefore the obtained extremum is indeed a maximum. Also, because the maximum is very narrow, we can approximate its curvature by the parabolic descent and find that the width of the maximum is of the order of

$$\delta D \propto \sqrt{\frac{P_s(E\varepsilon)(1-P_s(E\varepsilon))}{N}}. \tag{8}$$

Fluctuations of $D$ in equilibrium have an order of the maximum width and therefore relative fluctuations are inversely proportional to the square root of the number of fibers $N$: $\frac{\delta D}{D^{equil}} \propto \frac{1}{\sqrt{N}}$. Indeed, the logarithm of the probability $W_{macro\,D}^{equil}(D)$ of fluctuations in the vicinity of the maximum can be approximated by the parabolic dependence $\ln W_{macro\,D}^{equil}(D) = \ln W_{macro\,D}^{equil}(D_0) + \frac{1}{2}\frac{d^2 \ln W_{macro\,D}^{equil}}{dD^2}(D_0)\cdot \Delta D^2$ or

$W_{macro\,D}^{equil}(D) \propto \exp\left(-\frac{\Delta D^2}{2P_s(E\varepsilon)(1-P_s(E\varepsilon))/N}\right)$ and fluctuations are distributed in accordance with the Gaussian distribution. Standard deviation of this distribution is inversely proportional to the square root of $N$ and indeed for the relative fluctuations (ASM) we have $\frac{\sqrt{<\Delta D^2>}}{<D>} \propto \sqrt{\frac{1-P_s(E\varepsilon)}{P_s(E\varepsilon)}}\frac{1}{\sqrt{N}}$. So, the relative fluctuations of $D$ are inversely proportional to the square root of the number of fibers $N$ ($N$ is infinite in the thermodynamic limit). Therefore the maximum is indeed very narrow.

Ideally, to obtain any quantity in equilibrium, we must average it over all microstates: $f^{equil} \equiv \sum_n w_n^{equil} f_n = \sum_D N_D^{macro} w^{equil}(D) f(D)$. The fact that the maximum is very narrow gives us a possibility (ASM) to calculate all quantities averaged over $D$ as their values at the point of the maximum: $f^{equil} \approx f(D_0)$. For example, the averaged damage parameter in the equilibrium equals to its value at the maximum: $D^{equil} \approx D_0$.

Is it possible for fluctuations to overwhelm the damping factor $\frac{1}{\sqrt{N}}$ and to stop to be Gaussian? This would be an indication of presence of a critical point. The formula $\frac{\sqrt{<\Delta D^2>}}{<D>} \propto \sqrt{\frac{1-D_0}{D_0}}\frac{1}{\sqrt{N}}$ contains only a trivial singularity $D_0 = 0$. Similar, for the fraction of intact fibers $L \equiv 1-D$ the relative fluctuation $\frac{\sqrt{<\Delta L^2>}}{<L>} \propto \sqrt{\frac{D_0}{1-D_0}}\frac{1}{\sqrt{N}}$

contains the opposite trivial singularity $D_0 = 1$. The absence of non-trivial singularities suggests that the system under the EBC of constant strain does not have a critical point and fluctuations are always Gaussian. As we will see later this statement is not valid for the EBC of constant stress.

For the equilibrium probability distribution $w_n^{equil}$ the entropy is

$$S^{equil} \equiv -\sum_n w_n^{equil} \ln w_n^{equil} = -\sum_D N_D^{macro} w^{equil}(D) \ln w^{equil}(D). \tag{9}$$

Substituting Eq. (2a) we get

$$S^{equil} = -\sum_D N_D^{macro} w^{equil}(D)\{(1-D)N\ln(1-P_s(E\varepsilon)) + DN\ln P_s(E\varepsilon)\} = \tag{10}$$

$$= -(1-D^{equil})N\ln(1-P_s(E\varepsilon)) + D^{equil}N\ln P_s(E\varepsilon) = -\ln w_{D^{equil}} \approx -\ln w_{D_0}.$$

As $N_{D_0}^{macro} \approx_{\ln} 1/w(D_0)$ (ASM) $\tag{11}$

therefore $S^{equil} \approx \ln N_{D_0}^{macro}$. The number of microstates $\Delta\Gamma$ in the range of the width $\delta D$ of the maximum equals to the product of the number of microstates $N_D^{macro}$ for the given value of $D$ and the number of different $D$ in the region $\delta D$ of the maximum: $\Delta\Gamma \propto N_{D_0}^{macro} \dfrac{\delta D}{\Delta D}$. Here the width $\delta D$ of the maximum is given by Eq. (8) and $\Delta D = 1/N$ is the unit step of increment of $D$ due to the failure of one fiber (the unit step separating macrostates). Again, neglecting the power-law dependences on $N$ in comparison with the exponential dependence we obtain $\Delta\Gamma \approx_{\ln} N_{D_0}^{macro}$. Therefore

$$S^{equil} \approx \ln N_{D_0}^{macro} \approx \ln \Delta\Gamma \tag{12}$$

We have obtained the fundamental result (ASM): the definition of entropy as the negative logarithm of microstate probability averaged over all microstates $S \equiv -<\ln w_n>$ is equivalent to another definition of the entropy as the logarithm of the number of microstates over which the system presumably realizes itself. Using Eq. (3) for the entropy of equilibrium state (7) we obtain

$$S^{equil} \approx \ln N_{D_0}^{macro} \approx -NP_s(E\varepsilon)\ln P_s(E\varepsilon) - N(1-P_s(E\varepsilon))\ln(1-P_s(E\varepsilon)) \qquad (13)$$

It should be mentioned that the entropy given by Eqs. (9-10, 12-13) is the statistical, topological entropy and must not be confused with the thermodynamic entropy. In the literature it is often to multiply Eq. (12) by the Boltzmann constant $k_B$ and to use this quantity in the energy-balance equation as the thermodynamic entropy. The incorrectness of this approach can be easy illustrated by the following example. Let's assume that only one fiber is broken. In this case the entropy given by Eq. (12) and multiplied by $k_B$ is $S \approx k_B \ln N$. If the number of fibers is high but finite, e.g., $N = e^{10}$, the product of the entropy and temperature $TS$ has an order of the energy of 10 degrees of freedom. In contrast, the surface energy and elastic energy of the broken fiber as a macroscopic object have the order of the Avogadro constant ($10^{23}$ time higher). Therefore, it is irrelevant to use the obtained expression of entropy in the energy-balance equation because the entropy given by Eq. (12) does not represent the thermodynamic entropy. Instead, a new, 'topological' analog of the energy-balance equation must be constructed as we will see below.

# 6 Equation of state for the constant stress σ = const as an EBC

Let us first consider a simple example. For the EBC of constant strain we had $\sigma = (1-D_0)E\varepsilon = (1-P_s(E\varepsilon))E\varepsilon$ for the virtual stress at the model boundary. Its derivative is $\frac{d\sigma}{d\varepsilon} = -E^2\varepsilon p_s(E\varepsilon) + E(1-P_s(E\varepsilon))$. The value of $\varepsilon$ at which the derivative $\frac{d\sigma}{d\varepsilon}$ equals to zero is a fracture point separating regions of stable and unstable equilibriums for the EBC $\sigma$ = const. For the EBC $\varepsilon$ = const of course both regions are stable.

As an example let us obtain a solution for the uniform strength distribution

$$p_s(s) = \begin{cases} 0, & s < s_1 \\ \frac{1}{s_2 - s_1}, & s_1 < s < s_2 \\ 0, & s_2 < s \end{cases}, \quad P_s(s) = \begin{cases} 0, & s < s_1 \\ \frac{s - s_1}{s_2 - s_1}, & s_1 < s < s_2 \\ 1, & s_2 < s \end{cases}.$$

The equation of state can be easy obtained. For any values of parameters $s_1$ and $s_2$ the virtual stress $\sigma$ has a linear dependence on strain $\varepsilon$ in the range $0 < E\varepsilon < s_1$. This is the elastic behavior of the model till the first fiber failure. If $s_2 \leq 2s_1$ then after the strain $\varepsilon$ exceeds value $s_1/E$ the stress monotonically decreases with the further increase of the strain. Instability condition $\frac{d\sigma}{d\varepsilon} < 0$ means that for the EBC of constant stress the non-elastic branch of the stress curve is always unstable and a complete rupture follows the first fiber failure. However, if $s_2 > 2s_1$ then the stress with the further increase of strain increases initially, comes to the point where $\frac{d\sigma}{d\varepsilon} = 0$, and then decreases monotonically. This means that the initial part of the non-elastic curve is stable while its continuation after the point $\frac{d\sigma}{d\varepsilon} = 0$ is unstable and causes the complete rupture. The behavior of the model is illustrated by Fig. 1.

Point $\frac{d\sigma}{d\varepsilon}=0$ is a fracture point S of the model for the EBC $\sigma = $ const (see Fig. 3). For any values of model parameters each possible value of the virtual stress $\sigma$ has two strain solutions $\varepsilon_A$ and $\varepsilon_B$ (and respectively two damage solutions $D_A$ and $D_B$) as it is shown in Fig. 2. However, in the case of the constant strain Eq. (7) has only one solution as the direct dependence of $D$ on $\varepsilon$ ($D_A$ corresponds to $\varepsilon_A$ and $D_B$ corresponds to $\varepsilon_B$) and for any value of constant strain the probability $W^{equil}_{macro\,D}(D)$ has only one extremum (a maximum). For $N = 100$ and $\varepsilon = $ const an example of the dependence of $W^{equil}_{macro\,D}(D)$ on $D$ is given in Fig. 4a.

For the external condition $\sigma = $ const a solution can be obtained in a way similar to the previous section. Maximization of (4b) gives

$$\ln\left(\frac{P_s\left(\frac{\sigma}{1-D}\right)}{D}\frac{1-D}{1-P_s\left(\frac{\sigma}{1-D}\right)}\right) = \left\{\frac{1-D}{1-P_s\left(\frac{\sigma}{1-D}\right)} - \frac{D}{P_s\left(\frac{\sigma}{1-D}\right)}\right\}\frac{\sigma}{(1-D)^2}P'_s\left(\frac{\sigma}{1-D}\right) \quad (14)$$

where $P'_s(x)$ is the derivative of the function $P_s(x)$. This equation has a solution similar to Eq. (7):

$$D_0 = P_s\left(\frac{\sigma}{1-D_0}\right) \quad (15)$$

for the equilibrium state. However, now two different values of the damage parameter ($D_A$ and $D_B$ in Fig. 2) correspond to Eq. (15) for any value of the constant external stress. Therefore the dependence $W^{equil}_{macro\,D}(D)$ has two equal local maxima as it is shown in Fig. 4b for the example above. Also there is a third, non-trivial solution for the

Eq. (14) which however gives not a maximum of $W^{equil}_{macro\,D}(D)$ but a minimum between two maxima (15).

The forth solution is a point F (Fig. 3) of a complete fracture $D_F = 1$. Indeed, for this case we have $N^{macro}_{D_F} \equiv 1$, $w^{equil}(D_F) = \left\{ P_s\left(\dfrac{\sigma}{1-D_F}\right) \right\}^{D_F N} \equiv 1$ and $W^{equil}_{macro\,D_F}(D_F) = 1$. All four extremum values of $D$ are given in Fig. 4b.

Behavior of fluctuations in this case requires more careful investigation. Differentiating $\ln W^{equil}_{macro\,D}(D)$ in the vicinity of the equilibrium (15) we have

$$\ln W^{equil}_{macro\,D}(D) = \ln W^{equil}_{macro\,D}(D_0) - \frac{1}{2}\frac{N}{D_0(1-D_0)}\left\{ 1 - \frac{\sigma}{(1-D_0)^2} P'_s\left(\frac{\sigma}{1-D_0}\right) \right\} \cdot \Delta D^2 + ... \qquad (16)$$

Differential of Eq. (15) gives $dD_0 = \left\{ \dfrac{d\sigma}{1-D_0} + \dfrac{\sigma}{(1-D_0)^2} dD_0 \right\} P'_s\left(\dfrac{\sigma}{1-D_0}\right)$. Then for the fluctuations $\Delta D$ around the maximum of $W^{equil}_{macro\,D}(D)$ we obtain

$$W^{equil}_{macro\,D}(D) \propto \exp\left\{ -\frac{\Delta D^2}{\dfrac{1}{N} D_0\left(1 - D_0 + \sigma \dfrac{dD_0}{d\sigma}\right)} \right\} \qquad (17)$$

where $\dfrac{dD_0}{d\sigma}$ is the derivative corresponding to the change of the equilibrium damage $D_0$ with the equilibrium change of the constant external stress $\sigma$. For the relative fluctuations of the damage parameter we obtain

$$\frac{\sqrt{<\Delta D^2>}}{<D>} \propto \frac{1}{\sqrt{N}} \sqrt{\frac{1}{D_0}\left(1 - D_0 + \sigma \frac{dD_0}{d\sigma}\right)}. \qquad (18)$$

Now we already have the factor $\sqrt{\left(1-D_0+\sigma\frac{dD_0}{d\sigma}\right)}$ capable to overwhelm the $\frac{1}{\sqrt{N}}$ attenuation. Indeed, this expression has a singularity when $\frac{dD_0}{d\sigma}=\infty$ or $\frac{d\sigma}{dD_0}=0$. Remembering that $E\varepsilon=\frac{\sigma}{1-D_0}$, for the point of singularity we obtain $\frac{d\sigma}{d\varepsilon}=0$. This corresponds to the point S in Fig. 3 and as it was suggested by many authors [7, 9, 14] is *in situ* a spinodal point of the model. As the constant external stress $\sigma$ approaches this point $\sigma|_S$ the minimum of $W_{macro\,D}^{equil}(D)$ between two solutions (15) becomes shallower and at the point S two maxima of $W_{macro\,D}^{equil}(D)$ coalesce. At this value of the external stress the second derivative of $W_{macro\,D}^{equil}(D)$ in the maximum becomes zero and the behavior of fluctuations is determined already by the non-zero fourth derivative

$$\ln W_{macro\,D}^{equil}(D)=\ln W_{macro\,D}^{equil}(D_0)-\frac{1}{8}\frac{N}{\frac{E\varepsilon}{\sigma}-1}\frac{E^4\varepsilon^8}{\sigma^6}\left(\frac{d^2\sigma}{d\varepsilon^2}\right)^2\cdot\Delta D^4+...$$

The typical behavior of $W_{macro\,D}^{equil}(D)$ for this case is given in Fig. 4c. The fluctuations are no more Gaussian and the relative fluctuations become proportional to $1/\sqrt[4]{N}$. This behavior is typical for the spinodal point in the meanfield Landau's theory (ASM) and its applicability to the FBM model seems to be natural because the model we discuss is the meanfield model. For the case of a FBM with the local stress redistribution in the vicinity of the point S the size of fluctuations in the system would have power-law divergence which would correspond to the infinite correlation length (ASM). Therefore, under the EBC of the constant stress the system has a spinodal point and the damage in the system exhibits properties of the first order phase transition A → B → F.

We have found three independent equilibrium values of damage parameter $D_A$, $D_B$, and $D_F$ for the dependence $W_{macro\,D}^{equil}(D)$. And in accordance with our solution all three values should be stable. However, we know that the equilibrium $D_A$ is *in situ* metastable and that the equilibrium $D_B$ is *in situ* unstable because $\frac{d\sigma}{d\varepsilon}<0$ there. It is again in the complete analogy with the Van der Waals' meanfield model for the gas-liquid system. The exact solution of the homogeneous meanfield Van der Waals' model also gives the metastable and unstable branches of the isothermal pressure-volume dependence. The reason is that the constructed model is *a priori* homogeneous. Only introduction of the Maxwell's rule for heterogeneous liquid-gas transition allows substituting the unstable branch by the stable coexistence of two phases. The case of the FBM is similar. Our model is *a priori* meanfield homogeneous model. Therefore its exact solution exhibits the presence of the unstable branch. Only introduction of heterogeneity into the system would allow substituting this branch by the stable Maxwell's solution. If we were to imagine a heterogeneous FBM with local range of stress transfer we would have a heterogeneous mixture of intact and broken states A and F. It is illustrated in Fig. 3. Curve 0 - S is metastable, curve S – B - G is unstable, and the Maxwell's rule is given by the straight line A - B - F. Here point F represents the infinity of the strain therefore actually curve A – B - F should be thought as a horizontal line to infinity.

So, we have found that under the EBC of constant stress the FBM exhibits presence of the first order phase transition and spinodal. However, as it will be shown further, the classification of phase transitions is controversial because the continuity of the free energy potential strongly depends on the choice of the free energy potential itself. Some potentials, although they acquire a minimum at equilibrium, can be continuous even in the case of the first order phase transitions. This will be illustrated by Eq. (24) in the next paragraph.

**7 Temperature and free energy potential**

To construct the free energy potential it is necessary to discuss first the derivation of the free energy potential in SM. First we consider a thermodynamic system isolated with the given energy $E$. In the energy spectrum of the system some $g(E)$ degenerated levels correspond to this value of energy $E$. As to degenerated levels we will for simplicity refer to the groups of levels with close values of energies. For the case of an ideal system without interactions it is possible to think of the exact degeneracy in the sense of quantum mechanics.

So, only $g(E)$ microstates are possible for the isolated system. The equilibrium probability of each microstate is $w_n^{equil} = 1/g(E)$. The entropy of the system in equilibrium is

$$S^{equil} \equiv -\sum_{n=1}^{g(E)} w_n^{equil} \ln w_n^{equil} = -g(E)\frac{1}{g(E)} \ln \frac{1}{g(E)} = \ln g(E). \tag{19}$$

Here $g(E)$ is the number of microstates $\Delta\Gamma$ over which the system can realize itself with non-zero probability and $S^{equil} = \ln \Delta\Gamma$.

For the isolated system we have to use another definition of a macrostate: We define a non-equilibrium macrostate as a system that can realize itself only on a subset

$\Delta g$ of all possible $g(E)$ microstates: $\Delta g \subset g(E)$. Then the probability of each microstate for this macrostate is $w_n = 1/\Delta g$. There is already no superscript 'equil' in the probability here because this probability is not in equilibrium with the EBC $E =$ const. The entropy of the macrostate is $S^{macro}_{\Delta g} \equiv -\sum_{n=1}^{\Delta g} w_n \ln w_n = -\Delta g \frac{1}{\Delta g} \ln \frac{1}{\Delta g} = \ln \Delta g$. The probability of this macrostate in the isolated system (the probability to occur in equilibrium with the EBC $E =$ const) is $W^{equil}_{macro\,\Delta g} = \Delta g \cdot w^{equil}_n = \Delta g / g(E)$. This probability is *in situ* the free energy potential that should be maximized. The maximum of $W^{equil}_{macro\,\Delta g}$ corresponds to the equilibrium macrostate that occupies all possible microstates: $\Delta g_0 = g(E)$. The entropy of this macrostate equals the entropy of the system at the equilibrium: $S^{macro}_{\Delta g_0 = g(E)} = \ln g(E) = S^{equil}$.

Instead of the actual free energy potential $W^{equil}_{macro\,\Delta g}$ that should be maximized we can construct a potential that should be minimized. One of possible choices is $\Phi_{\Delta g} = -\ln W^{equil}_{macro\,\Delta g}$ because minus logarithm is a monotonically decreasing function. So defined potential identically equals to zero at the equilibrium macrostate $\Delta g_0 = g(E)$ as $W^{equil}_{macro\,\Delta g_0 = g(E)} = 1$. Another possible choice is $\Phi_{\Delta g} = -\ln(g(E) W^{equil}_{macro\,\Delta g})$ because $g(E)$ is a constant which does not influence the behavior of the potential. Now $\Phi_{\Delta g} = -\ln \Delta g = -S^{macro}_{\Delta g}$. Therefore the negative entropy of the non-equilibrium macrostates plays for the isolated system the role of the free energy potential that should be minimized.

For the FBM, instead of a system isolated with energy $E$ we can image a system isolated with the particular damage $D$. This system can realize only $N_D^{macro}$ microstates given by Eq. (3). Probability of each of these microstates is $w^{equil}(D) = 1/N_{macro\,D}$ (equilibrium with the isolation EBC) and the entropy of the system in equilibrium is

$$S^{equil} \equiv -\sum_{n=1}^{N_D^{macro}} w_n^{equil} \ln w_n^{equil} = -N_D^{macro} \frac{1}{N_D^{macro}} \ln \frac{1}{N_D^{macro}} = \ln N_D^{macro}.$$

For the non-equilibrium macrostate $\Delta N$ we have to use here an alternative definition, different from used in section 3. The non-equilibrium macrostate $\Delta N$ is defined as a macrostate when the system with probabilities $w_n = 1/\Delta N$ can realize itself only on the $\Delta N$ of all possible $N_D^{macro}$ microstates. The entropy of this macrostate is

$$S_{\Delta N}^{macro} \equiv -\sum_{n=1}^{\Delta N} w_n \ln w_n = -\Delta N \frac{1}{\Delta N} \ln \frac{1}{\Delta N} = \ln \Delta N$$

and the probability of this macrostate $\Delta N$ in the isolated system is $W_{macro\,\Delta N}^{equil} = \Delta N / N_D^{macro}$. This very probability $W_{macro\,\Delta N}^{equil}$ is the free energy potential that has to be maximized. Instead, we can construct the free energy potential that has to be minimized as $\Phi_{\Delta N} = -\ln W_{macro\,\Delta N}^{equil}$ or $\Phi_{\Delta N} = -\ln\left(N_D^{macro} W_{macro\,\Delta N}^{equil}\right) = -\ln \Delta N = -S_{\Delta N}^{macro}$. Again, ***the negative entropy*** of the non-equilibrium macrostates can be chosen as ***the free energy potential***.

Of course, our systems are non-Hamiltonian and the concept of energy is not applicable to them. Therefore we would suggest in future to use the term 'ruling' or 'governing' potential instead of 'free energy' potential. Or, being consistent with Sinai-Bowen-Ruelle's terminology [13, 15, 16], the term 'topological' or 'stochastic' potential could be used. However, to be consistent the term 'free energy' potential will be used till the end of the paper.

Phenomenological approach for an isolated system claims that the entropy of the isolated system can only increase: $\frac{dS}{dt} \geq 0$. We see that it corresponds to the fact that on its way to the equilibrium the system prefers macrostates with higher probability $W_{macro\,\Delta N}^{equil}$ (with higher $\Delta N$):

$$\frac{dS_{\Delta N}^{macro}}{dt} \geq 0. \qquad (20)$$

Now we consider the case of the canonical ensemble in SM ($N =$ const, $V =$ const, $T =$ const). In the canonical ensemble the temperature of the external medium as an ESC dictates to the system the equilibrium energy $E_0$ and the equilibrium probability of microstates

$$w_n^{equil} \equiv w^{equil}(E) = \frac{1}{z} e^{-\frac{E_n}{T}} \qquad (21)$$

where $z$ is the partition function of the system $z = \sum_n e^{-\frac{E_n}{T}}$. The entropy of the system in equilibrium is $S^{equil} \equiv -\sum_n w_n^{equil} \ln w_n^{equil} = -\sum_E g(E) w^{equil}(E) \ln w^{equil}(E)$ where the sum over microstates has been substituted by the sum over the values of energy and $g(E)$ denotes again the degeneration of the energy level $E$. Substituting Eq. (21) we obtain

$$S^{equil} = -\sum_E g(E) w^{equil}(E) \left\{ -\ln z - \frac{E}{T} \right\} = \ln z + \frac{E^{equil}}{T}. \qquad (22)$$

Defining the Helmholtz energy $A$ as $A \equiv E - TS$ (both for equilibrium and non-equilibrium states) we obtain $A^{equil} = E^{equil} - TS^{equil} = -T \ln z$. So, the Helmholtz energy

equals $-T\ln z$ only for equilibrium states. The equilibrium probability (21) is

$$w_n^{equil} = e^{\frac{A^{equil}-E_n}{T}}.$$

We can define a non-equilibrium macrostate as a subset of all microstates corresponding to the given energy $E$ (*i.e.*, as a system isolated with $E$). The number of these microstates is $g(E)$ and their probabilities are $w(E)=1/g(E)$ (for the system is constrained by this macrostate, *i.e.* isolated with the given $E$). The entropy of this macrostate is

$$S_E^{macro} = -g(E)\frac{1}{g(E)}\ln\frac{1}{g(E)} = \ln g(E) \qquad (23)$$

and the probability of this macrostate in the canonical ensemble (to occur in equilibrium with the EBC $T$ = const) is $W_{macro\,E}^{equil}(E) = g(E)w^{equil}(E) = g(E)e^{\frac{A^{equil}-E}{T}}$. This very probability function $W_{macro\,E}^{equil}(E)$ is the free energy potential that should be maximized. Also we can define the free energy potential that has to be minimized as $\Phi(E) = -\ln W_{macro\,E}^{equil}(E)$. The maximum of $W_{macro\,E}^{equil}(E)$ is very narrow, the number of energy levels $\Delta\Gamma$ in its width has an order of the degeneration of one of them $g(E_0)$ (again neglecting power-law dependences on $N$ in comparison with the exponential dependence of $g(E)$). But the area under the function $W_{macro\,E}^{equil}(E)$ has to accumulate its unity value under the maximum. Therefore we can conclude that at the maximum $g(E_0) \approx_{\ln} 1/w^{equil}(E_0)$ where $E_0$ is the equilibrium value of energy at the maximum. Therefore at the equilibrium macrostate

$$\Phi(E_0) = -\ln W_{macro\,E_0}^{equil}(E_0) = -\ln g(E_0) w^{equil}(E_0) \approx_{\ln} 0. \qquad (24)$$

For any equilibrium state this potential identically equals to zero. Therefore its derivatives over the equilibrium changes also equal to zero identically. Next we will introduce another free energy potential whose derivatives could be non-zero. Therefore the criterion to distinguish first and continuous phase transitions strongly depends on the choice of the free energy potential.

Also we can define the free energy potential that has to be minimized as $\Phi(E) = -\lambda_1 \ln\left(W_{macro\,E}^{equil}(E)\lambda_2\right)$ where $\lambda_1$ and $\lambda_2$ are some positive constants. Choosing these constants to be $\lambda_1 = T$ and $\lambda_2 = z$ we obtain

$$\Phi(E) = -T \ln W_{macro\,E}^{equil}(E) + A^{equil} = -T \ln g(E) - T \ln e^{\frac{A^{equil}-E}{T}} + A^{equil} = \qquad (25)$$

$$= -T \ln z_E^{macro} = E - T S_E^{macro} \equiv A_E^{macro}$$

where $z_E^{macro} \equiv \sum_{n:E_n=E} e^{-\frac{E_n}{T}} = g(E) e^{-\frac{E}{T}}$ is the partial partition function only over microstates corresponding to the given macrostate $E$. Free energy potential (25) corresponds to the Helmholtz energy for equilibrium and non-equilibrium states. While potential (24) is identically equal to zero for any equilibrium state and therefore its equilibrium derivatives are zero too, the Helmholtz energy (25) could have complex behavior of its derivatives characterizing the order of the possible phase transition in the system. Therefore the classification of orders of phase transitions significantly depends on the choice of the free energy potential.

The maximum is very narrow and $\ln w^{equil}(E)$ is a slowly changing function with a power-law dependence on $N$ in comparison with $g(E)$ and $w^{equil}(E)$ with the exponential dependence on $N$. Therefore for the entropy of the system in equilibrium we have

$$S^{equil} = -\sum_E g(E) w^{equil}(E) \ln w^{equil}(E) \approx -\ln w^{equil}(E_0) \sum_E g(E) w^{equil}(E) = \qquad (26)$$

$$= -\ln w^{equil}(E_0) \approx \ln g(E_0) \approx \ln \Delta\Gamma.$$

The entropy of the equilibrium macrostate is $S_{E_0}^{macro} = \ln g(E_0)$ and therefore again the entropy of the system at the equilibrium equals the entropy of the most probable macrostate and equals the logarithm of the number of microstates over which the system can realize itself.

At the maximum of $W_{macro\,E}^{equil}(E)$ we have $\dfrac{dW_{macro\,E}^{equil}}{dE}(E_0) = 0$ or $\dfrac{d\ln W_{macro\,E}^{equil}}{dE}(E_0) = 0$. For $\ln W_{macro\,E}^{equil}(E) = \ln g(E) + \dfrac{A^{equil} - E}{T}$ we obtain

$$\frac{1}{T} = \left.\frac{d\ln g(E)}{dE}\right|_{E_0} = \frac{d\ln g(E_0)}{dE_0} \qquad (27)$$

at the equilibrium state. Often this equation is used as a definition of temperature. As both the entropy of a macrostate $S_E^{macro} = \ln g(E)$ and the equilibrium entropy $S^{equil} \approx \ln g(E_0)$ have the same functional dependence on $E$ and $E_0$ respectively we obtain $\dfrac{1}{T} = \left.\dfrac{dS_E^{macro}}{dE}\right|_{E_0} \approx \dfrac{dS^{equil}}{dE_0}$. This is the energy-balance equation $dE_0 = TdS^{equil}$. For

non-equilibrium states instead, the increment of energy equals to the amount of heat received by the system $dE = \delta Q^{\leftarrow}$ where in general $\delta Q^{\leftarrow} < TdS$.

Imagine now a system in the canonical ensemble during its evolution over non-equilibrium macrostates $E$ on its way to the equilibrium. Each macrostate $E$ could be thought as a system isolated with the energy $E$. Therefore, the increase of the entropy in the system in accordance with Eq. (20) must be higher than the increase of the entropy produced only by the change of macrostates (23)

$$dS^{macro}_{\Delta N(E)} \geq dS^{macro}_{E} = d \ln g(E). \tag{28}$$

In the vicinity of the maximum of $W^{equil}_{macro\, E}(E)$ we have Eq. (27) and $dS \geq d \ln g(E) = dE/T$ or

$$dA \equiv d(E - TS) = dE - TdS \leq 0. \tag{29}$$

Therefore, we have confirmed that the Helmholtz energy is the free energy potential in the case of the canonical ensemble.

The behavior of the FBM under the EBC $\varepsilon$ = const is analogous to the behavior of the canonical ensemble and the damage parameter $D$ plays a role of the energy $E$. Indeed, the equilibrium probability of microstates assigned *a priori* by Eq. (2a) equals

$$w^{equil}(D) = \exp\bigl((1-D)N\ln(1 - P_s(E\varepsilon)) + DN\ln(P_s(E\varepsilon))\bigr) = \frac{1}{z}e^{-\frac{DN}{T}} \tag{30}$$

where $z = (1 - P_s(E\varepsilon))^{-N}$. It is easy to verify that $z$ is again the partition function of the system $z = \sum_D N_D^{macro} e^{-\frac{DN}{T}}$. The role of **the temperature T** prescribed by the ESC $P_s(E\varepsilon)$ is played by the quantity

$$T = \ln^{-1} \frac{1 - P_s(E\varepsilon)}{P_s(E\varepsilon)}. \tag{31}$$

The temperature here is a constant for the external constraint $\varepsilon$ = const and therefore could be used as an EBC too. Actually, two EBCs $\varepsilon$ = const or $T$ = const are equivalent and could be used intermittently.

Of course, the term 'temperature' here has nothing to do with the energy characteristics of the system and reflects only the similarity of statistics with SM. Therefore we would suggest naming it 'topological' temperature.

The entropy of the system in equilibrium is $S^{equil} \equiv -\sum_n w_n^{equil} \ln w_n^{equil} = -\sum_D N_D^{macro} w^{equil}(D) \ln w^{equil}(D)$ where the sum over microstates has been substituted by the sum over the values of damage and $N_D^{macro}$ is given by Eq. (3). Substituting Eq. (30) we obtain

$$S^{equil} = -\sum_D N_D^{macro} w^{equil}(D) \left\{ -\ln z - \frac{DN}{T} \right\} = \ln z + \frac{D^{equil} N}{T}. \tag{32}$$

Defining the Helmholtz energy $A$ as $A \equiv DN - TS$ we obtain $A^{equil} = D^{equil} N - TS^{equil}$ = $-T \ln z$. Therefore the Helmholtz energy equals $-T \ln z$ only for the equilibrium states. Then the equilibrium probability (30) is $w^{equil}(D) = e^{\frac{A^{equil} - DN}{T}}$.

We can define a non-equilibrium macrostate as a subset of all microstates corresponding to the given damage $D$ (*i.e.*, as a system isolated with $D$). The number of these microstates is $N_D^{macro}$ and their probabilities are $w(D) = 1/N_D^{macro}$. The entropy of this macrostate is

$$S_D^{macro} = -N_D^{macro} \frac{1}{N_D^{macro}} \ln \frac{1}{N_D^{macro}} = \ln N_D^{macro} \tag{33}$$

and the probability for this macrostate in equilibrium with the ESC is $W_{macro\,D}^{equil}(D) = N_D^{macro} w^{equil}(D) = N_D^{macro} e^{\frac{A^{equil} - DN}{T}}$. This very probability function $W_{macro\,D}^{equil}(D)$ is the free energy potential that should be maximized. Also we can define the free energy potential that has to be minimized as $\Phi(D) = -\ln W_{macro\,D}^{equil}(D)$. The maximum of $W_{macro\,D}^{equil}(D)$ is very narrow, the number $\Delta\Gamma$ of microstates in its range has an order of the number of microstates $N_{D_0}$ corresponding to one particular $D$ in this range (again neglecting power-law dependences on $N$ in comparison with the exponential dependence of $N_D^{macro}$). But the area under the function $W_{macro\,D}^{equil}(D)$ has to accumulate its unity value under the maximum. Therefore we can conclude that at the maximum $N_{D_0}^{macro} \approx_{\ln} 1/w^{equil}(D_0)$ where $D_0$ is the equilibrium value of damage given by Eq. (7). Also this result could be verified directly. Therefore at the equilibrium macrostate $\Phi(D_0) = -\ln W_{macro\,D_0}^{equil}(D_0) = -\ln N_{D_0}^{macro} w^{equil}(D_0) \approx 0$.

Also we can define the free energy potential that has to be minimized as $\Phi(D) = -\lambda_1 \ln\left(W_{macro\,D}^{equil}(D)\lambda_2\right)$ where $\lambda_1$ and $\lambda_2$ are some positive constants. Choosing these constants to be $\lambda_1 = T$ and $\lambda_2 = z$ we obtain

$$\Phi(D) = -T \ln W_{macro\,D}^{equil}(D) + A^{equil} = -T \ln N_D^{macro} - T \ln e^{\frac{A^{equil} - DN}{T}} + A^{equil} = \quad (34)$$

$$= -T \ln z_D^{macro} = DN - T S_D^{macro} \equiv A_D^{macro}$$

where $z_D^{macro} \equiv \sum_{n:D} e^{-\frac{DN}{T}} = N_D^{macro} e^{-\frac{DN}{T}}$ is the partial partition function only over microstates corresponding to the given macrostate $D$. Therefore now **the free energy potential** corresponds to **the Helmholtz energy** for equilibrium and non-equilibrium states.

As the maximum is very narrow for the entropy of the system at equilibrium we have

$$S^{equil} = -\sum_D N_D^{macro} w^{equil}(D) \ln w^{equil}(D) \approx -\ln w^{equil}(D_0) \sum_D N_D^{macro} w^{equil}(D) = \quad (35)$$

$$= -\ln w^{equil}(D_0) \approx \ln N_{D_0}^{macro} \approx \ln \Delta\Gamma.$$

The entropy of the equilibrium macrostate is $S_{D_0}^{macro} = \ln N_{D_0}^{macro}$. Therefore again the entropy of the system in the equilibrium equals the entropy of the most probable macrostate and equals the logarithm of the number of microstates over which the system can realize itself.

At the maximum of $W_{macro\,D}^{equil}(D)$ we have $\dfrac{dW_{macro\,D}^{equil}}{dD}(D_0) = 0$ or $\dfrac{d \ln W_{macro\,D}^{equil}}{dD}(D_0) = 0$.

For $\ln W_{macro\,D}^{equil}(D) = \ln N_D^{macro} + \dfrac{A^{equil} - DN}{T}$ we can write that

$$\frac{N}{T} = \left.\frac{d \ln N_D^{macro}}{dD}\right|_{D_0} = \frac{d \ln N_{D_0}^{macro}}{dD_0} \quad (36)$$

at the equilibrium state. This equation could be used as a definition of the temperature. As both the entropy of a macrostate $S_D^{macro} = \ln N_D^{macro}$ and the equilibrium entropy $S^{equil} \approx \ln N_{D_0}^{macro}$ have the same functional dependence on $D$ and $D_0$ respectively we obtain $\dfrac{N}{T} = \dfrac{dS_D^{macro}}{dD}\bigg|_{D_0} \approx \dfrac{dS^{equil}}{dD_0}$. This is an analog of the energy-balance equation - **the equation of stochastic, 'topological' balance** $NdD_0 = TdS^{equil}$. This equation could be obtained directly by the differentiating Eq. (33) as the logarithm of Eq. (3).

For non-equilibrium states instead, the increment of entropy is $dS \geq d\ln N_D^{macro} = NdD \cdot \ln\dfrac{1-D}{D}$. It is easy to see that for the equilibrium increment of entropy this formula gives the previous equation $dS^{equil} = NdD_0 \cdot \ln\dfrac{1-D_0}{D_0} = \dfrac{NdD_0}{T}$.

Imagine now a system in the canonical ensemble during its evolution over non-equilibrium macrostates $D$ on its way to the equilibrium. Each macrostate $D$ could be thought as a system isolated with $D$. Therefore the increase of the entropy in the system in accordance with Eq. (20) must be higher than the increase of the entropy produced only by the change of macrostates (33)

$$dS_{\Delta N(D)}^{macro} \geq dS_D^{macro} = d\ln N_D^{macro}. \tag{37}$$

In the vicinity of the maximum of $W_{macro\,D}^{equil}(D)$ we have Eq. (36) and

$$dS \geq d\ln N_D^{macro} = \dfrac{NdD}{T} \text{ or}$$

$$dA \equiv d(DN - TS) = NdD - TdS \leq 0.$$

So, we have confirmed that the Helmholtz energy is the free energy potential for the FBM under the external constraint $\varepsilon$ = const.

The behavior of the FBM under the EBC $\sigma$ = const is more complex. Indeed, the equilibrium probability of microstates is assigned *a priori* by Eq. (2b) and equals

$$w^{equil}(D) = \exp\left((1-D)N\ln\left[1-P_s\left(\frac{\sigma}{1-D}\right)\right] + DN\ln\left[P_s\left(\frac{\sigma}{1-D}\right)\right]\right). \tag{38}$$

This probability distribution is not a Gibbs-Boltzmann measure and therefore here we cannot develop an analogy with the canonical ensemble.

The entropy of the system in equilibrium is $S^{equil} \equiv -\sum_n w_n^{equil} \ln w_n^{equil} = -\sum_D N_D^{macro} w^{equil}(D) \ln w^{equil}(D)$ where the sum over microstates has been substituted by the sum over the values of damage, and $N_D^{macro}$ is given by Eq. (3). Substituting Eq. (38) we obtain

$$S^{equil} = -\sum_D N_D^{macro} w^{equil}(D)\left\{(1-D)N\ln\left(1-P_s\left(\frac{\sigma}{1-D}\right)\right) + DN\ln\left(P_s\left(\frac{\sigma}{1-D}\right)\right)\right\} = \tag{39}$$

$$= -(1-D_0)N\ln\left(1-P_s\left(\frac{\sigma}{1-D_0}\right)\right) - D_0 N\ln\left(P_s\left(\frac{\sigma}{1-D_0}\right)\right) = -(1-D_0)N\ln(1-D_0) - D_0 N\ln(D_0).$$

We have used here the fact that the maximum is very narrow and therefore average of any quantity equals to the value of this quantity at the point of the maximum.

We can again define a non-equilibrium macrostate as a subset of all microstates corresponding to the given $D$. The number of these microstates is $N_D^{macro}$ and their probabilities are $w(D) = 1/N_D^{macro}$. The entropy of this macrostate is

$$S_D^{macro} = -N_D^{macro} \frac{1}{N_D^{macro}} \ln \frac{1}{N_D^{macro}} = \ln N_D^{macro} \qquad (40)$$

and the probability of this macrostate in equilibrium with the ESC is

$$W_{macro\,D}^{equil}(D) = N_D^{macro} w^{equil}(D) = N_D^{macro} \exp\left((1-D)N \ln\left[1 - P_s\left(\frac{\sigma}{1-D}\right)\right] + DN \ln\left[P_s\left(\frac{\sigma}{1-D}\right)\right]\right).$$

This very probability function $W_{macro\,D}^{equil}(D)$ is the free energy potential that should be maximized. Also we can define a free energy potential that has to be minimized as $\Phi(D) = -\ln W_{macro\,D}^{equil}(D)$. The maximum of $W_{macro\,D}^{equil}(D)$ is very narrow, the number $\Delta\Gamma$ of microstates in its range has an order of the number of microstates $N_{D_0}^{macro}$ corresponding to one particular $D$ in this range (again neglecting power-law dependences on $N$ in comparison with the exponential dependence of $N_D^{macro}$). But the area under the function $W_{macro\,D}^{equil}(D)$ has to accumulate its unity value under the maximum. Therefore we can conclude that at the maximum $N_{D_0}^{macro} \approx_{\ln} 1/w^{equil}(D_0)$ where $D_0$ is the equilibrium value of damage given by Eq. (15). Also this result could be verified directly. Therefore at the equilibrium macrostate $\Phi(D_0) = -\ln W_{macro\,D}^{equil}(D_0) = -\ln N_{D_0}^{macro} w^{equil}(D_0) \approx 0$.

In accordance with the ASM we could expect that the free energy potential in the case of the constant external stress $\sigma$ would be a Gibbs potential. However, the temperature defined as $T = \ln^{-1} \frac{1 - P_s\left(\frac{\sigma}{1-D}\right)}{P_s\left(\frac{\sigma}{1-D}\right)}$ is no more a constant ESC of the system. This gives us a hint that the Gibbs potential is probably not a free energy potential in this case. Indeed, the true free energy potential $\Phi(D) = -\ln W_{macro\,D}^{equil}(D)$ can be written as

$$\Phi(D) = \frac{DN}{T} - \ln N_D^{macro} - N\ln\left(1 - P_s\left(\frac{\sigma}{1-D}\right)\right) = \frac{DN}{T} - S_D^{macro} - N\ln\left(1 - P_s\left(\frac{\sigma}{1-D}\right)\right) \qquad (41)$$

We see here the similar construction $\frac{DN}{T} - S_D^{macro}$ which is the Helmholtz energy $A_D^{macro}$ divided by temperature $T$. But in addition to this quantity we see also the term $-N\ln\left(1 - P_s\left(\frac{\sigma}{1-D}\right)\right)$ which has more complex dependence than the expected $\sigma\varepsilon \propto \frac{const}{1-D}$ for the Gibbs potential. It is the result of the fact that in the case of the constant external stress the behavior of the system is no more Gibbsian. In contrast, it obeys to the non-Gibbsian SM given by Eq. (38).

As the maximum is very narrow for the entropy of the system at equilibrium we have

$$S^{equil} = -\sum_D N_D^{macro} w^{equil}(D) \ln w^{equil}(D) \approx -\ln w^{equil}(D_0) \sum_D N_D^{macro} w^{equil}(D) = \qquad (42)$$

$$= -\ln w^{equil}(D_0) \approx \ln N_{D_0}^{macro} \approx \ln\Delta\Gamma.$$

The entropy of the equilibrium macrostate is $S_{D_0}^{macro} = \ln N_{D_0}^{macro}$. Therefore again the entropy of the system in the equilibrium equals the entropy of the most probable macrostate and equals the logarithm of the number of microstates over which the system can realize itself.

Differentiation of Eq. (3) gives

$$\frac{d\ln N_D^{macro}}{dD} \approx N\ln\left(\frac{1-D}{D}\right) \qquad (43)$$

and for the equilibrium state (15) we have

$$\left.\frac{d \ln N_D^{macro}}{dD}\right|_{D_0} \approx N \ln\left(\frac{1-D_0}{D_0}\right). \tag{44}$$

As both the entropy of a macrostate $S_D^{macro} = \ln N_D^{macro}$ and the equilibrium entropy $S^{equil} \approx \ln N_{D_0}^{macro}$ have the same functional dependence on $D$ and $D_0$ respectively we obtain $N \ln\left(\frac{1-D_0}{D_0}\right) \approx \left.\frac{dS_D^{macro}}{dD}\right|_{D_0} \approx \frac{dS^{equil}}{dD_0}$. This is **the equation of the topological balance** $NdD_0 = dS^{equil} \cdot \ln^{-1}\left(\frac{1-D_0}{D_0}\right)$. In fact, this equation is the direct differentiation of Eq. (40).

For non-equilibrium states the increment of entropy is $dS \geq d \ln N_D^{macro} = NdD \cdot \ln\frac{1-D}{D}$. So, we see that in general case the balance equation could be obtained by the differentiation of the non-equilibrium entropy. Peculiar here is the fact that for the general case the equation of topological balance depends actually only on the degeneration of macrostates $D$ given by Eq. (3) and does not depend on the equilibrium probability distribution given by Eq. (2). Therefore, only the structure of the system itself determines the balance equation and the ESC does not influence it. However, this constraint defines the equilibrium state and in this way ($1/T$ in Eqs. (27) and (36)) determines how the balance equation would look like in the equilibrium with this ESC.

So, for the general case the balance inequality is

$$dS \geq NdD \cdot \ln\frac{1-D}{D} \tag{45}$$

for equilibrium and non-equilibrium states. Again, Eq. (45) is valid for an arbitrary EBC, not only for the cases $\varepsilon = $ const or $\sigma = $ const. Substituting geometrical constraint (1) into Eq. (45) we obtain the balance inequality expressed in terms of the stress-strain ratio

$$dS \geq N \ln\left(\frac{E\varepsilon}{\sigma} - 1\right) d\left(\frac{\sigma}{E\varepsilon}\right) \tag{46}$$

This is the most general case of the balance inequality for the FBM with the fixed number of fibers.

## 9 Conclusion

An introduction of the model stochasticity as an external stochastic constraint (ESC) introduces fluctuating topological behavior into a system. For the FBM it is shown that these fluctuations 'statistically' thermalize in general a non-thermal system. The equilibrium 'canonical' distribution of probabilities is dictated by the ESC instead of the temperature of the external media. The formalism of the classical statistical mechanics is developed providing all classical features like the narrow probability maximum, free energy potential, balance equation, and equation of state. This gives rise to the new statistical mechanics, statistical mechanics of damage.

Behavior of the system exhibits the presence of a critical point and continuous phase transition in its vicinity. Apart from the critical point the system has the first order phase transition. However, the classification of the order of phase transition is shown to be controversial and is based on the choice of the free energy potential.

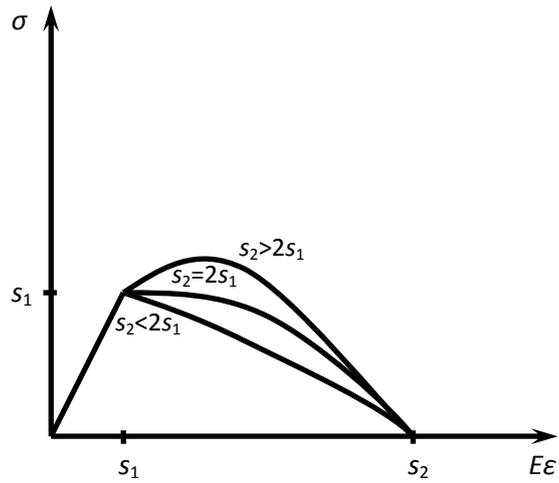

Fig. 1. The equation of state as the stress-strain dependence. Three curves illustrate the dependence for different ratios of $s_2/2s_1$.

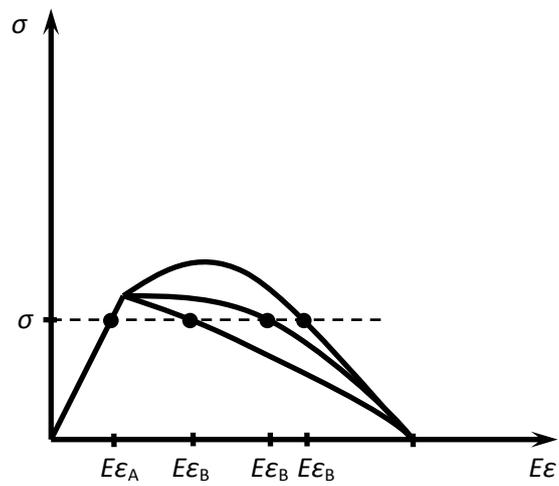

Fig. 2. Two solutions A and B for different ratios of $s_2/2s_1$.

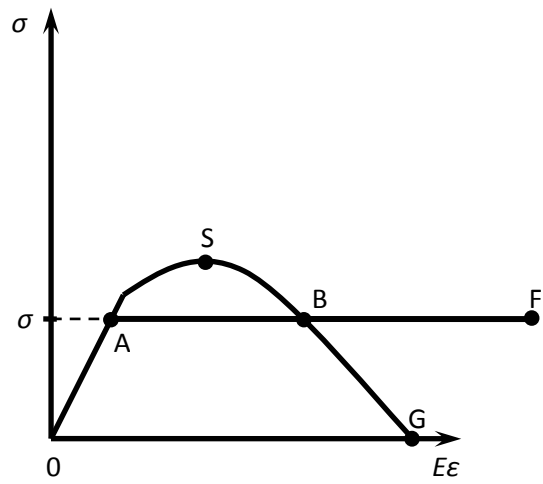

Fig. 3. Phase diagram. Point F is assumed to represent infinite strain. Straight line A - B - F corresponds to the Maxwell's rule.

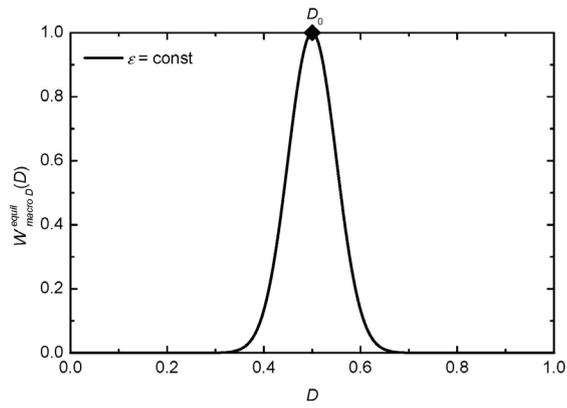
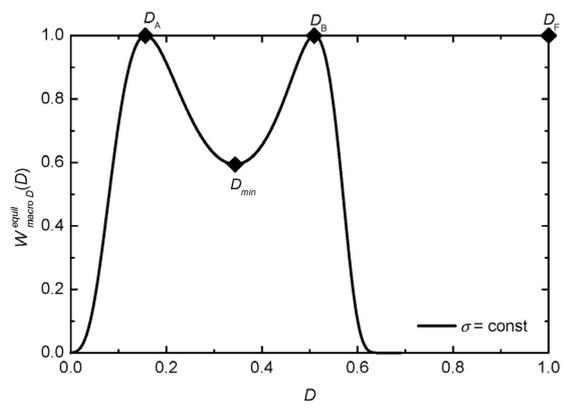

(a) (b)

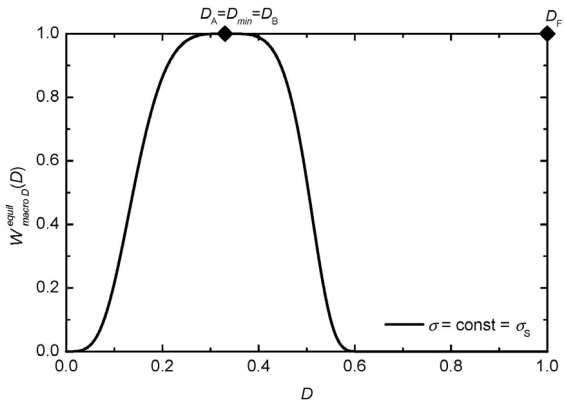

(c)

Fig. 4. Dependence of the probability $W_{macro\,D}^{equil}(D)$ on the non-equilibrium $D$ for the external constraint of (a) constant strain, (b) constant stress, and (c) constant stress at the spinodal point S.